# Flux-free growth of large superconducting crystal of FeSe by traveling-solvent floating-zone technique

**Mingwei Ma, Dongna Yuan, Yue Wu, Huaxue Zhou, Xiaoli Dong and Fang Zhou**

Beijing National Laboratory for Condensed Matter Physics, Institute of Physics, Chinese Academy of Sciences, Beijing 100190, China

Email: fzhou@aphy.iphy.ac.cn

**Abstract**

A flux-free solution to the growth of large and composition homogeneous superconducting FeSe crystal is reported for the first time, which is based on the traveling-solvent floating-zone technique. The size of the crystal samples prepared by this approach is up to $15\times6\times2$ mm$^3$, being far bigger than previously reported in all dimensions, and the main phase of the crystals is of a single preferred orientation along the tetragonal (101) plane. X-ray diffraction analysis identifies the main phase to be the superconducting tetragonal β-FeSe. The superconducting transition temperature ($T_C$) is determined to be 9.4 K by AC magnetic susceptibility and electronic transport measurements. A nearly perfect diamagnetic shielding of -97% is observed, indicating a bulk superconductivity in the crystal sample.

The discovery of iron-based superconductor La[$O_{1-x}F_x$]FeAs [1], the 1111-type, and a superconductivity up to 55 K in this family [2] in 2008 has triggered tremendous interest of high-$T_C$ superconductivity community. Subsequent works have reported the superconductivity in other iron-based materials such as (Ba/Sr/Ca,K/Na)$Fe_2As_2$ [3-6] (122-type), LiFeAs (111-type) [7-9] and $Ca_{1-x}La_xFeAs_2$ (112-type) [10]. All these materials consist of similar FeAs layers. As one subgroup of the Fe-based materials, the 122-type alkalis iron selenide superconductors, such as $KFe_2Se_2$ [11], contain instead the layers of FeSe as the superconducting blocks. A high-$T_C$ superconductivity above 40 K can also be achieved in $A_xFe_2Se_2$ [12] and $LiFeO_2Fe_2Se_2$ [13], by intercalating in between the FeSe layers the metals of A (Li, Na, Ca, Eu, Yb) at room temperature and the spacer layer of $LiFeO_2$, respectively. Most interestingly, the simplest binary compounds of both FeAs and FeSe are also available, but only the material of FeSe itself, the 11-type, superconducts at 8 K [14]. And its $T_C$ increases up to 15 K with the substitution of Te for Se [15-17]. Moreover, FeSe exhibits a strong pressure effect on its $T_C$, which rises to 36.7 K under a pressure of 8.9 GPa [18] and even well above 50 K in monolayer FeSe films [19-21]. As a basic superconducting unit, featuring both the simplest chemical composition and the similar layered structure and exhibiting the unusual properties, FeSe is therefore thought to be the most important prototype for investigating the underlying mechanism of superconductivity in the family of iron-based superconductors. With this motivation, great efforts have been made to prepare superconducting FeSe crystals. For example, the flux technique is widely applied to grow the crystals using fluxes such as NaCl-KCl [22,23], LiCl-CsCl [24], KCl [25-28] and KCl-$AlCl_3$ [29]. However, the as-grown crystals always exhibit a small size. Even with well optimized crystal growth conditions, the biggest superconducting FeSe crystals prepared from a KCl solution are only of 6×3×0.1~0.2 $mm^3$ or 5×5×0.1~0.2 $mm^3$ in dimensions [28]. The limitation in sample size has obstructed some important data collection on the crystals of FeSe, such as the data of inelastic neutron scattering. Generally speaking, a low solubility of the Fe/Se solutes in the flux solutions is the main cause for the small crystal size. It is therefore necessary to adopt some other techniques based on a melt, not a flux, growth to obtain larger crystals of FeSe, which are highly demanded by, for example, neutron scattering experiments. Actually, Yang et al. [30] have tried to grow FeSe crystals via a flux-free

Bridgman method. However, the as-prepared materials are polycrystals showing multiple preferred crystal orientations.

In this communication, we report, for the first time, another flux-free solution to the growth of large and composition homogeneous superconducting FeSe crystals, which is based on the traveling-solvent floating-zone (TSFZ) technique. The size of the FeSe crystals prepared by this approach is up to 15×6×2 mm$^3$, which is far bigger than that of the previous works, and the crystals exhibit a single preferred orientation of a tetragonal (101) plane for the superconducting main phase. Here the principle of this growth technique is sketched as follows. With the commonly used KCl flux method, the crystal growth of FeSe usually proceeds by a slow cooling from a soaking temperature of about 850℃ to about 770℃, the melting point of KCl (for example, the case of ref [28]). According to the binary phase diagram of Fe-Se system as displayed in fig. 1 [31], the superconducting tetragonal β-FeSe phase forms at a narrow region below 457 ℃, therefore the crystal directly grown at that temperature range is the non-superconducting δ-FeSe. At lower temperatures below 457 ºC, the most part of as-grown δ-FeSe is phase transformed into the superconducting β-FeSe. In the present work, however, the crystal growth is carried out at an elevated temperature of 990 ºC without any flux but using a solvent of the composition Fe:Se=1:0.87. As illustrated in the inset of fig. 1, the solid S (i.e. δ'-FeSe$_{0.96}$) melts incongruently at 990 ºC and is in equilibrium with the liquid L with Fe:Se=1:0.87. Note that the slope of the liquidus line near the point of L is not steep, which is beneficial for the crystallization of the liquid. The unique advantage of the traveling-solvent floating-zone technique lies in that the solid-liquid equilibrium between S and L can be maintained during a stable TSFZ growth process, so that a big and composition homogeneous crystal S can be prepared. And the superconducting β-FeSe is subsequently obtained due to the phase transformation at lower temperatures below 457 ºC.

High purity powders of Se (Alfa Aesar, 99.999%) and Fe (Sinopharm Chemical, 99.99%) were used as the raw materials. The initial composition of the feed rod for TSFZ growth was taken as that of the S point in fig. 1, with Fe:Se=1:0.96 in mole ratio, and that of the solvent the L point with Fe:Se=1:0.87. The accurately weighed raw materials for the feed rod and solvent were ground and mixed with an agate mortar and pestle in glove box filled with argon gas. The mixtures were then pressed into separate columnar shapes before sealed in an evacuated quartz tube and heated at 700℃ for 10 hours. The obtained products were once again well ground, sealed in evacuated quartz tube and heated at 700 ºC for 10 hours. After that, the materials were ground into fine powders in the glove box. The preparation of dense and homogeneous feed rods is one of the key factors in achieving a stable and continuous TSFZ growth. The fine powders of 8 g in mass used for the feed rod was loaded into a tubular balloon and pressed under a high hydrostatic pressure over 5 kbar for 2 minutes to form a compact cylindrical rod of 8 cm in length. The feed rod was then sealed in evacuated quartz tube and was finally sintered at 750℃ for 3 hours. The solvent used for the TSFZ growth was prepared through the same procedures as that of the feed rod and was finally pressed and sintered together with the feed rod.

An optical floating-zone furnace (Crystal Systems Inc., Model FZ-T-10000-H) with 4×150 W halogen lamps as infrared radiation sources was used for TSFZ growth. The crystal was grown in a highly purified argon flow at a zone traveling rate of 0.6 mm/h, using a previously grown FeSe crystal as a seed crystal which is oriented along the tetragonal (101) plane. During the crystal growth the upper feed rod and the lower grown ingot were rotated in opposite directions to improve growth conditions. The composition of as-prepared crystals was analyzed by inductively coupled plasma atomic emission spectroscopy (ICP-AES). Powder and single crystal x-ray diffraction measurements were carried out at room temperature on an x-ray diffractometer (MXP18A-HF) using Cu $K_\alpha$ radiation, with 2θ ranging from 10º to 80º and a 2θ scanning step of 0.01º. Experiments of x-ray rocking curve were performed on a double-crystal diffractometer (Delta-X HRXRD) equipped with a Ge (004) monochromator (α~10") under a power of 1.6 kW. The AC magnetic susceptibility was determined on Quantum Design MPMS XL-1 with a frequency of 997.34 Hz and an amplitude of 1 Oe in a temperature range from 2 K to 20 K. The in-plane resistance of the crystal sample was measured on Quantum Design PPMS-9 using the standard 4-probe method from room temperature down to 2 K.

Figure 2 shows the photos for an as-grown ingot with a size of 5–6 mm in diameter and 100 mm in length (in the upper) and for some typical FeSe crystals cleaved from as-grown ingots (in the lower). The biggest crystal is of about 15×6×2 mm$^3$ in dimensions. This crystal size is far larger than previously reported in all dimensions. The composition of the crystals is determined by ICP-AES to be FeSe$_{1\pm0.01}$, which is close to the initial composition of the feed rod. The composition of the crystals is very homogeneous as demonstrated in the lower right part of fig. 2, where the composition analyzed by ICP-AES is almost the same at four different locations along the biggest crystal of FeSe.

Displayed in fig. 3(a) is the powder XRD pattern of the crystal sample. All the main diffraction peaks can be indexed on a previously reported tetragonal structure of β-FeSe with the space group of P4/nmm. The lattice constants are calculated to be a=3.771Å, c=5.522Å and volume V=78.55Å$^3$. Besides the main reflections, some weak peaks from the second hexagonal δ-FeSe phase are also detected as what is reported previously [22-28]. Our experiments indicate that a long time annealing of FeSe crystal at 400 ºC for up to 10 days only reduce the hexagonal δ-FeSe phase to some extent, but can not remove it completely. Given in figure 3(b) is a typical XRD pattern obtained on the FeSe crystal. Except for a very weak peak from the second hexagonal phase, the two reflections of (h0h) type indicate that the main phase of the crystal is of a single preferred orientation along the tetragonal (101) plane. Actually, many XRD experiments are made to check the preferred crystal orientation on some typical crystals such as shown in fig.2, including the XRD checks on the reverse sides of them and on the surfaces exposed by cleaving thick crystal pieces. All those XRD patterns are similar to that of figure 3(b). Illustrated in fig. 4 is the x-ray rocking curve of the (101) Bragg reflection for a FeSe crystal having a surface dimensions of 5×5 mm$^2$ (with the beam slit size being set to 1×10 mm$^2$). The large full-width-at-half maximum (FWHM) of 5.3° signifies a considerable crystal mosaicity in this material. In fact, the x-ray rocking curve obtained on a FeSe$_{0.94}$ crystal grown from a KCl flux solution also gives the same result [28], implying that this crystal imperfection is related to some intrinsic crystallization course of FeSe. Nevertheless, the spin excitation spectrum of neutron scattering is not significantly affected by the large sample mosaic, based on the initial neutron experiment data collected on our crystals.

The superconductivity of the FeSe crystal is characterized by both AC magnetic susceptibility and electronic transport measurements, as shown in fig. 5. The superconducting transition temperature is determined to be 9.4 K from the temperature dependence of AC magnetic susceptibility corrected for demagnetization factor (fig. 5a), and the temperature dependence of in-plane resistance exhibits a zero resistance temperature of 9.3 K (inset of fig. 5b). The real part of AC susceptibility at 2 K is -0.97, this nearly perfect diamagnetic shielding indicates a bulk superconductivity in the crystal sample.

In conclusion, large and composition homogeneous superconducting FeSe crystals have been successfully grown for the first time via a new solution based on the traveling-solvent floating-zone technique without using any flux. The size of the crystal samples prepared by this approach is up to $15 \times 6 \times 2$ mm$^3$, being far bigger than previously reported in all dimensions, and the main phase of the crystal samples is of a single preferred orientation along the tetragonal (101) plane. XRD analysis identifies the main phase to be the superconducting tetragonal β-FeSe, with the lattice parameters being a=3.771Å and c=5.522Å. Double-crystal x-ray rocking curve shows a FWHM of 5.3° for the crystal. The considerable mosaicity of this material may be attributed to its intrinsic crystallization. The superconducting transition temperature is determined to be 9.4 K based on the AC magnetic susceptibility and electronic transport measurements. A nearly perfect diamagnetic shielding signal at 2 K is observed from the real part of the AC susceptibility, indicating a bulk superconductivity in the crystal.

**Acknowledgments**

We thank Dr. Lihong Yang for her technical assistance in XRD measurements, Mr. Hongjun Shi and Mr. Li Wang for their help in ICP-AES analysis. This work is supported by projects

**References**

[1] Kamihara Y, Watanabe T, Hirano M and Hosono H 2008 *J. Am. Chem. Soc.* **130** 3296

[2] Ren Z A *et al* 2008 *Chin. Phys. Lett.* **25** 2215

[3] Rotter M, Tegel M and Johrendt D 2008 *Phys. Rev. Lett*. **101** 107006

[4] Chen G F, Li Z, Li G, Hu W Z, Dong J, Zhou J, Zhang X D, Zheng P, Wang N L and Luo J L 2008 *Chin. Phys. Lett.* **25** 3403

[5] Sasmal K, Lv B, Lorenz B, Guloy A M, Chen F, Xue Y Y and Chu C W 2008 *Phys. Rev. Lett.* **101** 107007

[6] Wu G, Chen H, Wu T, Xie Y L, Yan Y J, Liu R H, Wang X F, Ying J J and Chen X H 2008 *J. Phys. Condens. Matter* **20** 422201

[7] Tapp J H, Tang Z J, Lv B, Sasmal K, Lorenz B, Chu P C W and Guloy A M 2008 *Phys. Rev. B* **78** 060505(R)

[8] Wang X C, Liu Q Q, Lv Y X, Gao W B, Yang L X, Yu R C, Li F Y and Jin C Q 2008 *Solid State Commun.* **148** 538

[9] Pitcher M J, Parker D R, Adamson P, Herkelrath S J C, Boothroyd A T, Ibberson R M, Brunelli M and Clarke S J 2008 *Chem. Commun.* **2008** 5918

[10] Katayama N *et al* 2013 *J. Phys. Soc. Jpn.* **82** 123702

[11] Guo J G, Jin S F, Wang G, Wang S C, Zhu K X, Zhou T T, He M and Chen X L 2010 *Phys. Rev. B* **82** 180520

[12] Ying T P, Chen X L, Wang G, Jin S F, Zhou T T, Lai X F, Zhang H, Wang W Y 2012 *Sci. Rep.* **2** 426

[13] Lu X F, Wang N Z, Zhang G H, Luo X G, Ma Z M, Lei B, Huang F Q and Chen X H

2013 *Phys. Rev. B* **89** 020507(R)

[14] Hsu F C *et al* 2008 *Proc. Natl. Acad. Sci. USA* **105** 14262

[15] Yeh K W *et al* 2008 *Eur. Phys. Lett.* **84** 37002

[16] Fang M H, Pham H M, Qian B, Liu T J, Vehstedt E K, Liu Y, Spinu L and Mao Z Q 2008 *Phys. Rev. B* **78** 224503

[17] Yeh K W, Hsu H C, Huang T W, Wu P M, Huang Y L, Chen T K, Luo J Y and Wu M K 2008 *J. Phys. Soc. Jpn.* **77** 19

[18] Medvedev S *et al* 2009 *Nature Mater.* **8** 630

[19] Wang Q Y *et al* 2012 *Chin. Phys. Lett.* **29** 037402

[20] He S L *et al* 2013 *Nature Mater.* **12** 605

[21] Tan S Y *et al* 2013 *Nature Mater.* **12** 634

[22] Zhang S B *et al* 2009 *Supercond. Sci. Technol.* **22** 015020

[23] Zhang S B *et al* 2009 *Supercond. Sci. Technol.* **22** 075016

[24] Hu R W, Lei H C, Abeykoon M, Bozin E S, Billinge S J L, Warren J B, Siegrist T and Petrovic C 2011 *Phys. Rev. B* **83** 224502

[25] Mok B H *et al* 2009 *Cryst. Growth Des.* **9** 3260

[26] Rao S M *et al* 2011 *J. Appl. Phys.* **110** 113919

[27] Gorina J I, Kaluzhnaya G A, Golubkov M V, Rodin V V, Sentjurina N N and Chernook S G 2012 *Cryst. Rep.* **57** 585

[28] Ma M W, Yuan D N, Wu Y, Dong X L and Zhou F 2014 *Physica C* http://dx.doi.org/10.1016/j.physc.2014.06.003.

[29] Chareev D, Osadchii E, Kuzmicheva T, Lin J Y, Kuzmichev S, Volkova O and Vasiliev A 2013 *CrystEngComm.* **15** 1989

[30] Yang C M, Chen P W, Kou J C, Diko P, Chen I G and Wu M K 2011 *IEEE Trans. Appl. Supercond.* **21** 2845

[31] Okamoto H 1991 *J. Phase Equilib.* **12** 383

Figure 1

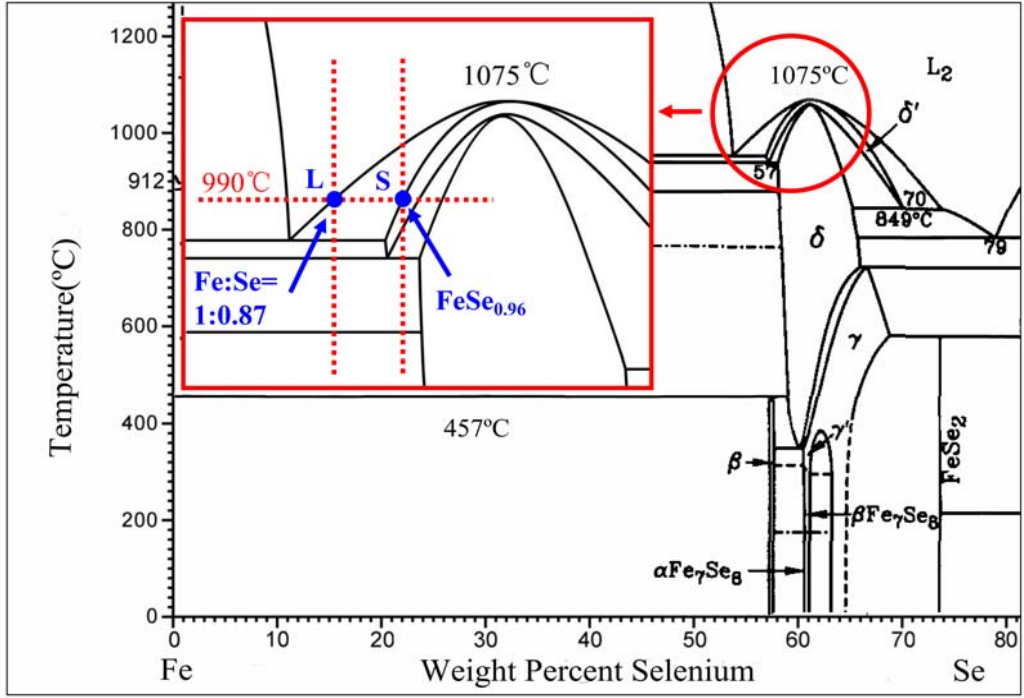

Figure 1. Fragment of Fe-Se phase diagram [31]. The superconducting tetragonal β-FeSe phase forms at the narrow region near 57.5 wt% of Se below 457 ℃. The inset is a magnified local area showing the solid-liquid equilibrium at 990 ºC between S (δ'-FeSe$_{0.96}$) on the solidus and L (Fe:Se=1:0.87) on the liquidus. See the text for details.

Figure 2

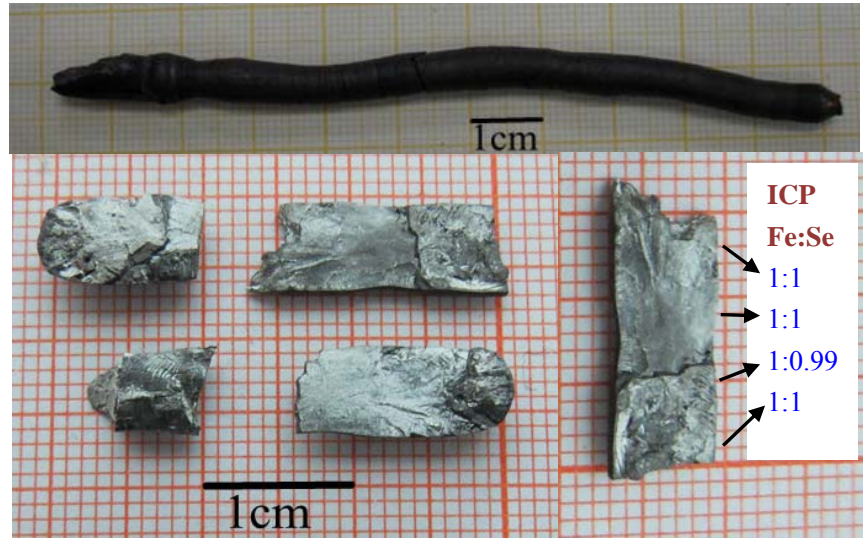

Figure 2. Photographs for an as-grown ingot with a size of 5–6 mm in diameter and 100 mm in length (the upper one) and for some typical FeSe crystal pieces cleaved from as-grown ingots (the lower ones). The biggest crystal is of about 15×6×2 mm$^3$ in dimensions. The composition of the crystals is very homogeneous as demonstrated in the lower right part of fig. 2, where the composition analyzed by ICP-AES is almost the same at four different locations along the biggest crystal of FeSe. The background mesh division is 1 mm.

Figure 3

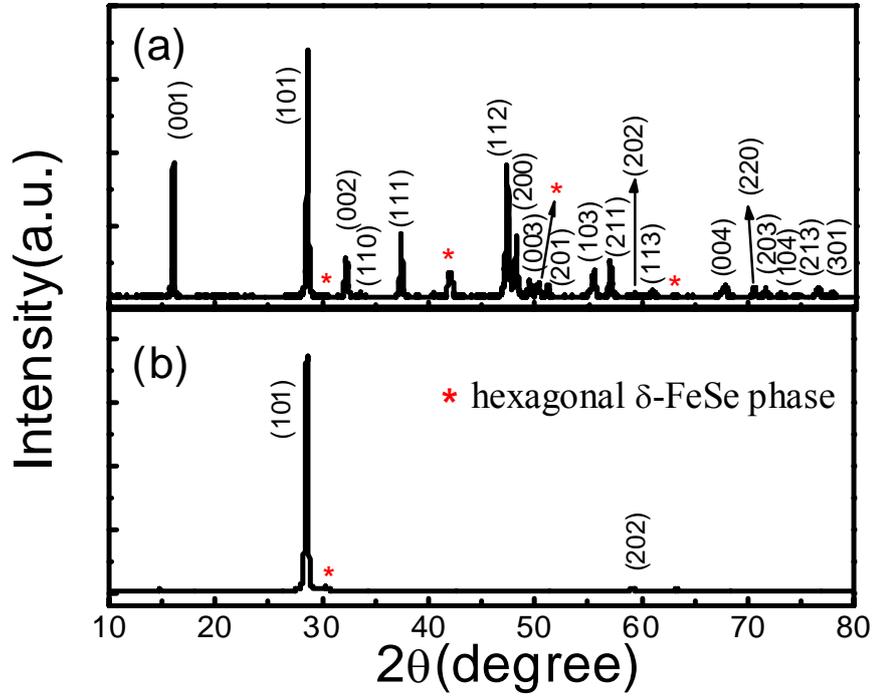

Figure 3. (a) Powder x-ray diffraction pattern of FeSe crystal. All the main diffraction peaks are indexed on a previously reported tetragonal structure of β-FeSe, with lattice parameters being a=3.771 Å and c=5.522 Å, except for some weak peaks from the second hexagonal phase as indicated by asterisks. (b) A typical XRD pattern obtained on the FeSe crystal. Again a very weak peak from the second hexagonal phase is detected.

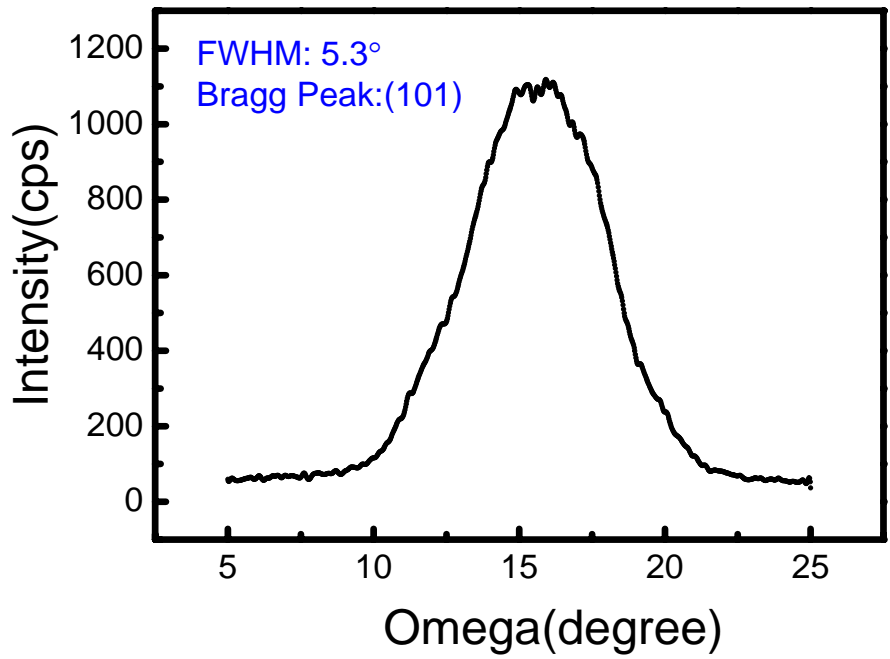

Figure 4. Rocking curve of the tetragonal (101) reflection for FeSe crystal taken by double-crystal x-ray diffraction using Cu Kα1 radiation. The surface dimension of the sample is 5×5 mm$^2$ and the beam slit size is 1×10 mm$^2$.

Figure 5

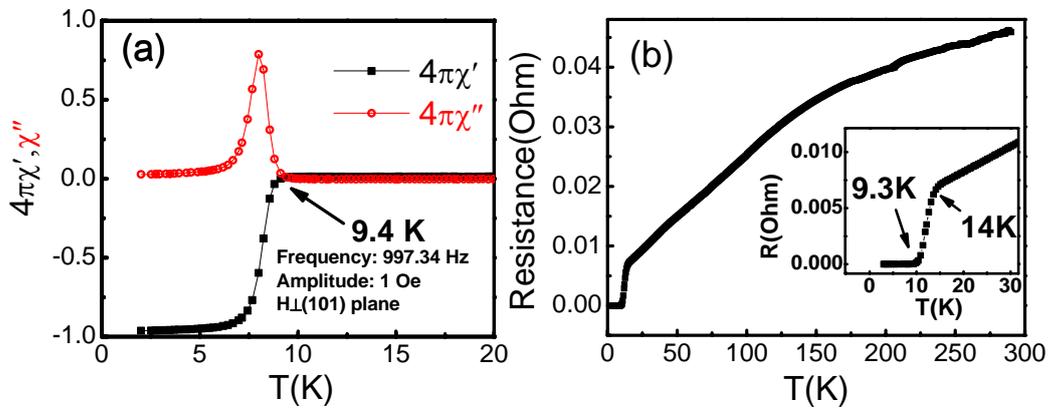

Figure 5. (a) Temperature dependence of AC magnetic susceptibility of the FeSe crystal corrected for demagnetization factor. The superconducting transition temperature $T_C$ is determined to be 9.4 K, and the diamagnetic shielding signal at 2 K to be -0.97 from the real part of AC susceptibility. (b) Temperature dependence of in-plane resistance of the FeSe crystal. The inset is a magnified plot of the data at lower temperatures, which shows a zero resistance temperature of 9.3 K.

**Figure Captions**

Figure 1. Fragment of Fe-Se phase diagram [31]. The superconducting tetragonal β-FeSe phase forms at the narrow region near 57.5 wt% of Se below 457 ℃. The inset is a magnified local area showing the solid-liquid equilibrium at 990 ºC between S ($\delta'$-FeSe$_{0.96}$) on the solidus and L (Fe:Se=1:0.87) on the liquidus. See the text for details.

Figure 2. Photographs for an as-grown ingot with a size of 5–6 mm in diameter and 100 mm in length (the upper one) and for some typical FeSe crystal pieces cleaved from as-grown ingots (the lower ones). The biggest crystal is of about 15×6×2 mm$^3$ in dimensions. The composition of the crystals is very homogeneous as demonstrated in the lower right part of fig. 2, where the composition analyzed by ICP-AES is almost the same at four different locations along the biggest crystal of FeSe. The background mesh division is 1 mm.

Figure 3. (a) Powder x-ray diffraction pattern of FeSe crystal. All the main diffraction peaks are indexed on a previously reported tetragonal structure of β-FeSe, with lattice parameters being a=3.771 Å and c=5.522 Å, except for some weak peaks from the second hexagonal phase as indicated by asterisks. (b) A typical XRD pattern obtained on the FeSe crystal. Again a very weak peak from the second hexagonal phase is detected.

Figure 4. Rocking curve of the tetragonal (101) reflection for FeSe crystal taken by double-crystal x-ray diffraction using Cu Kα1 radiation. The surface dimension of the sample is 5×5 mm$^2$ and the beam slit size is 1×10 mm$^2$.

Figure 5. (a) Temperature dependence of AC magnetic susceptibility of the FeSe crystal corrected for demagnetization factor. The superconducting transition temperature T$_C$ is determined to be 9.4 K, and the diamagnetic shielding signal at 2 K to be -0.97 from the real part of AC susceptibility. (b) Temperature dependence of in-plane resistance of the FeSe crystal. The inset is a magnified plot of the data at lower temperatures, which shows a zero resistance temperature of 9.3 K.